\documentclass[aps,prc,twoside,twocolumn,nofootinbib,
showpacs,floatfix,dvips]{revtex4}

\usepackage{amsmath,amssymb,graphicx,bm}

\begin{document}


\title{\boldmath Isospin-dependent pion in-medium effects on charged pion ratio in heavy ion collisions}


\author{Jun Xu}

\email{xujun@comp.tamu.edu}

\affiliation{Cyclotron Institute, Texas A\&M University, College Station, Texas 77843, U.S.A.}

\author{Che Ming Ko}

\email{ko@comp.tamu.edu}

\affiliation{Cyclotron Institute and
Department of Physics and Astronomy, Texas A\&M University, College
Station, Texas 77843, U.S.A.}

\author{Yongseok Oh}

\email{yoh@kisti.re.kr}

\affiliation{Korea Institute of Science and Technology Information, Daejeon 305-806, Korea}

\altaffiliation{Address after March 1, 2010: School of Physics and Energy Sciences, Kyungpook National University, Daegu 702-701, Korea}

\date{\today}

\begin{abstract}

Using results from the chiral perturbation theory for the $s$-wave
interaction and the delta-resonance model for the $p$-wave
interaction of pions with nucleons, we have evaluated the spectral
functions of pions in asymmetric nuclear matter with unequal proton
and neutron densities. We find that in hot dense neutron-rich
matter the strength of the spectral function of positively charged
pion at low energies is somewhat larger than that of negatively
charged pion.  In a thermal model, this isospin-dependent effect
slightly reduces the ratio of negatively charged to positively
charged pions that are produced in heavy ion collisions induced by
radioactive beams. Relevance of our results to the determination of
the nuclear symmetry energy from measured ratio of negatively to
positively charged pions produced in heavy ion collisions is
discussed.
\end{abstract}

\pacs{21.65.Jk, 
      21.65.Ef, 
      25.75.Dw  
      }

\maketitle

\section{Introduction}

The nuclear symmetry energy is the energy needed per nucleon to
convert all protons in a symmetric nuclear matter to neutrons.
Knowledge on the density dependence of the nuclear symmetry energy
is important for understanding the dynamics of heavy ion collisions
induced by radioactive beams, the structure of exotic nuclei with
large neutron or proton excess, and many important issues in nuclear
astrophysics~\cite{LCK08,ireview98,ibook,baran05}. At normal nuclear
matter density, the nuclear symmetry energy has long been known to
have a value of about $30$~MeV from fitting the binding energies of
atomic nuclei with the liquid-drop mass formula. Somewhat stringent
constraints on the nuclear symmetry energy below the normal nuclear
density have also been obtained during past few years from studies
of the isospin diffusion~\cite{Tsa04,Liu07,Che05a,LiBA05c} and
isoscaling~\cite{She07} in heavy-ion reactions, the size of neutron
skin in heavy nuclei~\cite{Ste05b}, and the isotope dependence of
the giant monopole resonances in even-A Sn isotopes~\cite{Gar07}.
For nuclear symmetry energy at high densities, transport model
studies have shown that the ratio of negatively to positively
charged pions produced in heavy ion collisions with neutron-rich
nuclei is sensitive to its
stiffness~\cite{bali02,Ferini:2006je,Ferrini:2005jw}, and there has
been an attempt \cite{Xiao09} to extract the symmetry energy at
supra-saturation densities from the FOPI data on the $\pi ^{-}/\pi
^{+}$ ratio in central heavy-ion collisions at SIS/GSI \cite{FOPI}.

The transport model used in Ref.~\cite{bali02,Ferini:2006je,Xiao09}
neglects, however, medium effects on pions, although it includes
those on nucleons and produced $\Delta$ resonances through their
isospin-dependent mean-field potentials and scattering cross
sections. It is well-known that pions interact strongly in nuclear
medium as a result of their $p$-wave interactions through couplings
to the nucleon-particle--nucleon-hole and
delta-particle--nucleon-hole ($\Delta$-hole) excitations, leading to
the softening of their dispersion relations or increased strength of
their spectral functions at low
energies~\cite{weise75,friedmann81,oset82,xia94,hees05,korpa08}.
Furthermore, it has been shown in Ref.~\cite{korpa99} that in
asymmetric nuclear matter with different proton and neutron
fractions, these medium effects depend on the charge of the pion.
Including the $p$-wave pion medium effects in the transport model
has previously been shown to enhance the production of low energy
pions in high energy heavy ion collisions, although it does not
affect the total pion yield~\cite{xiong93,Fuchs:1996pa}.  Properties
of pions in nuclear medium are also modified by their $s$-wave
interactions with nucleons. Studies based on chiral perturbation
theory have shown that in asymmetric nuclear matter this effect also
depends on the charge of the pion~\cite{Kaiser:2001bx}, leading to
different in-medium masses for negatively and positively charged
pions.  Since pions of different charges are modified differently in
asymmetric nuclear matter, it is of interest to study how such
isospin-dependent medium effects would affect the ratio of negatively to positively
charged pions produced in heavy ion collisions.

This paper is organized as follows. We first evaluate in
Sec.~\ref{medium} the spectral functions of pions and
delta-resonances in hot dense asymmetric nuclear matter.
Using these spectral functions, we  then
study in Sec.~\ref{pionratio} the charged pion ratio in a thermal
model and compare the results with available experimental data. We
conclude with a summary in Sec.~\ref{summary}.

\section{Medium effects}\label{medium}

\subsection{Pion in-medium properties}

Pions in nuclear medium acquire self-energies through their $s-$wave
and $p$-wave interactions. In asymmetric nuclear matter, the pion
self energy further depends on the charge of the pion.

\subsubsection{Pion-nucleon $s$-wave interaction}

The contribution of the pion-nucleon $s$-wave interaction
to the pion self-energy has been calculated in Ref.~\cite{Kaiser:2001bx} up to the two-loop order
in chiral perturbation theory. In asymmetric nuclear matter
of proton density $\rho_p$ and neutron density $\rho_n$, the resulting pion self-energy
depends on the charge of the pion, and for $\pi^-$, $\pi^+$, and
$\pi^0$, their self-energies due to the $s$-wave interactions are given, respectively, by
\begin{eqnarray}
\Pi_s^-(\rho_p,\rho_n)&=&\rho_n[T^-_{\pi N}-T^+_{\pi N}]-\rho_p[T^-_{\pi N}+T^+_{\pi N}]
\notag\\
&&+\Pi^-_{\rm rel}(\rho_p,\rho_n)+\Pi^-_{\rm cor}(\rho_p,\rho_n)\notag\\
\Pi_s^+(\rho_p,\rho_n)&=&\Pi_s^-(\rho_n,\rho_p)\notag\\
\Pi_s^0(\rho_p,\rho_n)&=&-(\rho_p+\rho_n)T^+_{\pi N}+\Pi^0_{\rm cor}(\rho_p,\rho_n).
\end{eqnarray}
In the above, $T^\pm$ are the isospin-even and isospin-odd $\pi
N$ $s$-wave scattering $T$-matrices, which are given by the one-loop
contribution in chiral perturbation theory and have the empirical values $T^-_{\rm \pi N}\approx
1.847~{\rm fm}$ and $T^+_{\rm \pi N}\approx -0.045~{\rm fm}$
extracted from the energy shift and width of the 1s level in pionic
hydrogen atom. The term $\Pi^-_{\rm rel}$ is due to the relativistic correction,
whereas the terms $\Pi^-_{\rm cor}$, and $\Pi^0_{\rm cor}$ are the
contributions from the two-loop order in chiral perturbation theory.
Their expressions can be found in Ref.~\cite{Kaiser:2001bx}, where
it is also shown that in nuclear matter of density $\rho=0.165~{\rm fm}^{-3}$ and isospin asymmetry $\delta=(\rho_n^{} - \rho_p^{})/(\rho_n^{} + \rho_p^{})=0.2$, such as
in the interior of a Pb nucleus, changes of the pion
masses due to their $s$-wave interactions are $\Delta
m_{\pi^-}=13.8~{\rm MeV}$, $\Delta m_{\pi^+}=-1.2~{\rm MeV}$, and
$\Delta m_{\pi^0}=6.1~{\rm MeV}$. Compared to their masses in free space,
the $\pi^-$ mass  increases whereas the $\pi^+$ mass
decreases in neutron-rich nuclear matter.

\subsubsection{Pion-nucleon $p$-wave interaction}

For the $p$-wave interactions of pions in nuclear matter,
we consider only the dominant $\Delta$-hole excitations as in
Ref.~\cite{ko89}, as the contribution from the nucleon particle-hole excitations
is known to be small.  For a pion of isospin state $m_t$,
energy $\omega$, and momentum $k$ in a hot nuclear medium
at temperature $T$, its self-energy due to the $p$-wave interaction is then given by
\begin{eqnarray}\label{pi1}
\Pi_0^{m_t} &\approx& \frac{4}{3}
\left(\frac{f_\Delta^{}}{m_\pi^{}} \right)^2 k^2 F_\pi^2(k)
\sum_{m_\tau,m_T^{}} \left|\left\langle {\textstyle\frac{3}{2}}\, m_T^{} | 1\, m_t\, {\textstyle\frac{1}{2}} \, m_\tau \right\rangle \right|^2\notag\\
&\times& \int \frac{d^3p}{(2\pi)^3}\frac{1}{e^{(m_N^{} + p^2/2m_N^{}
+U_N^{m_\tau}-\mu_B^{}-2m_\tau\mu_Q^{})/T}+1}
\notag\\
&\times&
\left(\frac{1}{\omega-\omega_{m_T^{}}^{+}}+\frac{1}{-\omega-\omega_{m_T^{}}^{-}}\right),
\end{eqnarray}
with
\begin{eqnarray}
\omega_{m_T^{}}^{\pm} &\approx& m_\Delta^{} +U_\Delta^{m_T^{}} +
\frac{(\vec{k} \pm \vec{p})^2}{2m_\Delta^{}} -i\frac{\Gamma_\Delta^{m_T^{}}}{2}\notag\\
&-&m_N^{}-U_N^{m_\tau} - \frac{p^2}{2 m_N^{}}. \notag
\end{eqnarray}
In the above, $m_\Delta^{} \simeq 1232$~MeV is the mass of $\Delta$
resonance; $f_\Delta^{}\simeq 3.5$ is the $\pi N\Delta$ coupling
constant and $F_\pi(k) = [1+0.6(k^2/m^2_\pi)]^{-1/2}$~\cite{art} is
the $\pi N\Delta$ form factor determined by fitting the decay width
$\simeq 118$~MeV of $\Delta$ resonance in free space. The summation
in Eq.~(\ref{pi1}) is over the nucleon isospin state $m_\tau$ and
the $\Delta$ resonance isospin state $m_T^{}$; and the factor
$\langle {\textstyle\frac{3}{2}} \, m_T^{} | 1\,
m_t\,{\textstyle\frac{1}{2}} \, m_\tau \rangle$ is the
Clebsch-Gordan coefficient from the isospin coupling of pion with
nucleon and $\Delta$ resonance. The momentum integration is over
that of nucleons in the nuclear matter given by a Fermi-Dirac
distribution with $\mu_B$ and $\mu_Q$ being, respectively, the
baryon and charge chemical potentials determined by charge and
baryon number conservations; $U_N^{m_\tau}$ is the mean-field
potential of nucleons of isospin state $m_\tau$ in nuclear matter;
and $\Gamma_\Delta^{m_T^{}}$ and $U_\Delta^{m_T^{}}$ are,
respectively, the width and mean-field potential of $\Delta$
resonance of isospin state $m_T^{}$.

\begin{figure}[h]
\centerline{\includegraphics[width=3.2in,height=3.2in,angle=0]{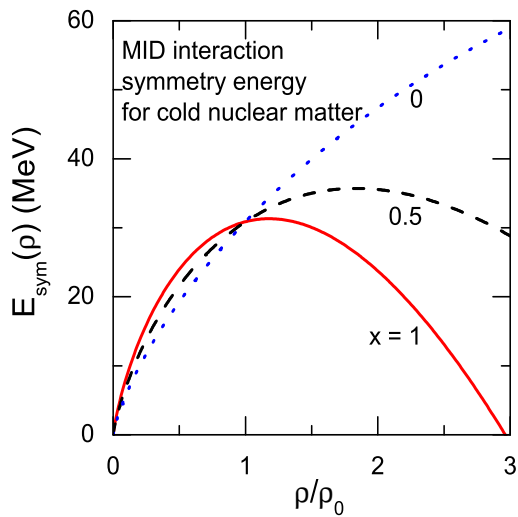}}
\caption{(Color online) Nuclear symmetry energies as functions of
nuclear density from the MID interaction for different values of the
symmetry energy parameter $x$.}\label{symmetry}
\end{figure}

For the nucleon mean-field potential $U_N^{m_\tau}$, we use that
from the momentum-independent (MID) interaction~\cite{LCK08}, i.e.,
\begin{eqnarray}
U_N^{m_\tau}(\rho_B^{},\delta_{\rm like}) =&& \alpha\left(\frac{\rho_B^{}}{\rho_0^{}}\right) + \beta\left(\frac{\rho_B^{}}{\rho_0^{}}\right)^\gamma\notag\\
&+&U_{\text{asy}}^{m_\tau}(\rho_B^{} ,\delta_{\rm like}),
\end{eqnarray}
with
\begin{eqnarray}
&&U_{\text{asy}}^{m_\tau}(\rho_B^{} ,\delta_{\rm like})=\notag\\
&&-4\left\{ F(x)\left(\frac{\rho_B^{}}{\rho_0^{}}\right) + [18.6-F(x)] \left(\frac{\rho_B^{}}{\rho_0^{}}\right)^{G(x)}\right\} m_\tau \delta_{\rm like} \notag\\
&&+[18.6 - F(x)][G(x) - 1] \left(\frac{\rho_B^{}}{\rho_0^{}}\right)^{G(x)} {\delta_{\rm like} }^{2}
\end{eqnarray}
being the nucleon symmetry potential. The parameters
$\alpha=-293.4$~MeV, $\beta=240.1$~MeV, and $\gamma=1.216$ are
chosen to give a compressibility of $212$~MeV and a binding energy
per nucleon of $-16$~MeV for symmetric nuclear matter at saturation
or normal nuclear density $\rho_0^{} = 0.16~{\rm fm}^{-3}$. The
nucleon symmetry potential $U_{\rm asy}^{m_\tau}(\rho_B^{} ,\delta_{\rm like})$
depends on the baryon density $\rho_B^{} = \rho_n^{} + \rho_p^{} +
\rho_{\Delta^-}^{} + \rho_{\Delta^0}^{} + \rho_{\Delta^+}^{} +
\rho_{\Delta^{++}}^{}$ and the isospin asymmetry $\delta_{\rm
like}=(\rho_n^{} - \rho_p^{} + \rho_{\Delta^-}^{} -
\rho_{\Delta^{++}}^{}+ \rho_{\Delta^0}^{}/3 -
\rho_{\Delta^+}^{}/3)/\rho_B^{}$ of asymmetric hadronic matter,
which is a generalization of the isospin asymmetry $\delta$ usually
defined for asymmetric nuclear matter without $\Delta$
resonances~\cite{bali02}. The nucleon mean-field potential also
depends on the stiffness of nuclear symmetry energy through the
parameter $x$ via the functions $F(x)$ and $G(x)$. We consider the
three cases of $x=0$, $x=0.5$, and $x=1$ with corresponding values
$F(x=0) = 129.98$ and $G(x=0) = 1.059$, $F(x=0.5) = 85.54$ and
$G(x=0.5) = 1.212$, and $F(x=1) = 107.23$ and $G(x=1) = 1.246$. The
resulting nuclear symmetry energy becomes increasingly softer as the
value of $x$ increases, with $x=1$ giving a nuclear symmetry energy
that becomes negative at about three times normal nuclear matter
density as shown in Fig.~\ref{symmetry}. These symmetry energies reflect the uncertainties in the
theoretical predictions on the stiffness of nuclear symmetry energy
at high densities.

For the mean-field potentials of $\Delta$
resonances, their isoscalar potentials are assumed to be the same as
those of nucleons, and their symmetry potentials are taken to be the
average of those for neutrons and protons with the weighting factors
depending on the charge state of $\Delta$ resonances~\cite{art},
i.e., $U_{\rm asy}^{\Delta^{++}} = U_{\rm asy}^p$, $U_{\rm
asy}^{\Delta^+} = {\textstyle\frac{2}{3}} U_{\rm asy}^p +
{\textstyle\frac{1}{3}} U_{\rm asy}^n$, $U_{\rm asy}^{\Delta^0} =
{\textstyle\frac{1}{3}} U_{\rm asy}^p + {\textstyle\frac{2}{3}}
U_{\rm asy}^n$, and $U_{\rm asy}^{\Delta^-} = U_{\rm asy}^n$.

Including the short-range $\Delta$-hole repulsive interaction via
the Migdal parameter $g^\prime$, which has values $1/3 \le g^\prime
\le 0.6$~\cite{weise75,friedmann81,oset82,xia94,hees05,korpa08},
modifies the pion self-energy due to the $p$-wave interaction to
\begin{eqnarray}
\Pi_p^{m_t} =\frac{\Pi_0^{m_t}}{1-g^\prime\Pi_0^{m_t}/k^2}.
\end{eqnarray}

\subsubsection{pion spectral function}

In terms of the pion self energies $\Pi_s^{m_\tau}$
and $\Pi_p^{m_\tau}$ due to the pion-nucleon $s$- and $p$-wave interactions
in nuclear medium, the pion in-medium propagator is given by
\begin{eqnarray}\label{pi2}
D^{m_t}(\omega,k) =
\frac{1}{\omega^2-k^2-m_\pi^2-\Pi_s^{m_t}-\Pi_p^{m_t}(\omega,k)}.
\end{eqnarray}
The pion spectral function $S_\pi^{m_t}(\omega,k)$ is related to the
imaginary part of the pion in-medium propagator through
\begin{eqnarray}
S_\pi^{m_t}(\omega,k) =
-\frac{1}{\pi}\,\mbox{Im}\,D^{m_t}(\omega,k).
\end{eqnarray}

\subsection{Delta resonance in-medium properties}

The modification of the pion properties in nuclear medium affects the
decay width and mass distribution of $ \Delta$ resonances. For a
$\Delta$ resonance of isospin state $m_T^{}$ and mass $M$ and at
rest in nuclear matter, its decay width is given by~\cite{ko89}
\begin{eqnarray}\label{gamma}
&&\Gamma_\Delta^{m_T^{}}(M)\approx -2 \sum_{m_\tau,m_t} |\langle {\textstyle\frac{3}{2}} \,
m_T^{} | 1\, m_t\, {\textstyle\frac{1}{2}} \, m_\tau \rangle|^2\notag\\
&\times& \int \frac{d^3{\bf k}}{(2\pi)^3} \left(\frac{f_\Delta^{}}{m_\pi^{}} \right)^2 F_\pi^2(k)\left[\frac{1}{z_\pi^{-1}e^{(\omega-m_t\mu_Q^{})/T}-1}+1\right]\notag\\
&\times&\left[1-\frac{1}{e^{(m_N^{} + k^2/2m_N^{} +U_N^{m_\tau}-\mu_B^{}-2m_\tau\mu_Q^{})/T}+1}\right]\notag\\
&\times&\mbox{Im}\, \left[\frac{k^2}{3}\frac{D^{m_t}(\omega,k)}{(1-g^\prime\Pi_0^{m_t}(\omega,k)/k^2)^2} +{g^\prime}^2\frac{\Pi_p^{m_t}(\omega,k)}{k^2} \right].\notag\\
\end{eqnarray}
In the above, the first term in the last line is due to the decay of
the $\Delta$ resonance to the pion but corrected by the contact
interaction at the $\pi N\Delta$ vertex, while the second term
contains the contribution from its decay to the $\Delta$-hole state
without coupling to the pion. The two temperature-dependent factors
in the momentum integral take into account, respectively, the Bose
enhancement for the pion and the Pauli blocking of the nucleon. To
include possible chemical non-equilibrium effect, a fugacity
parameter $z_\pi^{}$ is introduced for pions.
The pion energy $\omega$ is determined from energy conservation,
i.e., $M + U_\Delta^{m_T^{}} = \omega + m_N^{} + k^2/2m_N^{}
+U_N^{m_\tau}$. The resulting mass distribution of $\Delta$
resonances is then given by
\begin{eqnarray}
P_\Delta(M) = A\frac{\Gamma_\Delta^{m_T^{}}(M)/2}{(M-m_\Delta^{})^2 +{\Gamma_\Delta^{m_T^{}}}^2(M)/4},
\end{eqnarray}
where $A$ is a normalization constant to ensure the integration of $P_\Delta(M)$ over $M$ is one.

\subsection{Results}

\begin{figure}[t]
\centerline{\includegraphics[width=3.2in,height=3.2in,angle=0]{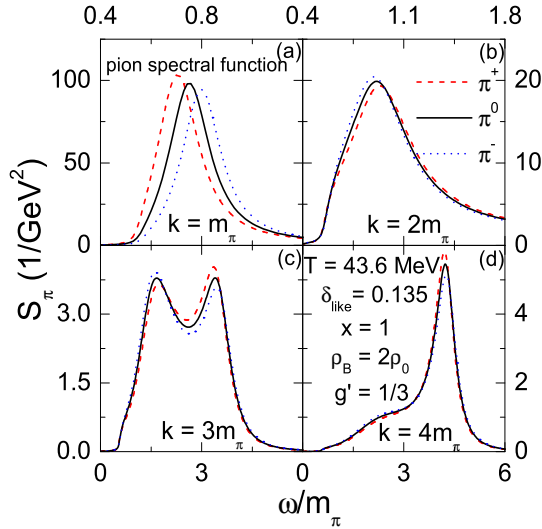}}
\caption{(Color online) Spectral functions of pions in asymmetric
nuclear matter of density $2\rho_0^{}$ and isospin asymmetry
$\delta_{\rm like}=0.135$ as functions of pion energy for different
pion momenta of (a) $m_\pi$, (b) $2 m_\pi$, (c) $3 m_\pi$, and (d)
$4 m_\pi$. All are calculated with the Migdal parameter
$g^\prime=1/3$.}\label{pion}
\end{figure}

\begin{figure}[h]
\centerline{\includegraphics[width=3.2in,height=3.2in,angle=0]{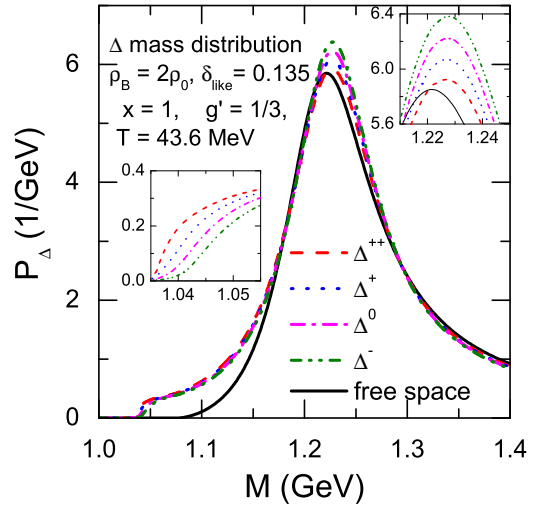}}
\caption{(Color online) Mass distributions of $\Delta$ resonances at
rest in asymmetric nuclear matter of density $2\rho_0^{}$ and
isospin asymmetry $\delta_{\rm like}=0.135$. The solid line
corresponds to that in free space. The distributions near the
threshold and at the peak are enlarged in the insets.} \label{delta}
\end{figure}

Equations~(\ref{pi1}), (\ref{pi2}) and (\ref{gamma}) are solved
self-consistently to obtain the pion spectral functions and the mass
distributions of $\Delta$ resonances in asymmetric nuclear matter.
The results obtained with the Migdal parameter $g^\prime = 1/3$ are
illustrated in Fig.~\ref{pion} and Fig.~\ref{delta} for an
asymmetric nuclear matter of isospin asymmetry $\delta_{\rm
like}\simeq 0.135$, twice the normal nuclear matter density
$\rho_B^{} = 2\rho_0^{}$, temperature $T\simeq 43.6~{\rm MeV}$, and
chemical potentials $\mu_B^{} \simeq 942~{\rm MeV}$ and $\mu_Q^{}
\simeq -18.5~{\rm MeV}$, corresponding to those to be used in the
thermal model and also similar to those reached in the IBUU
transport model with the nuclear symmetry energy $x=1$ for central
Au+Au collisions at the beam energy of $0.4~{\rm
AGeV}$~\cite{Xiao09}. Shown in Fig.~\ref{pion} are the pion spectral
functions as functions of pion energy for different values of the
pion momentum. It is seen that for low pion momenta the spectral
function at low energies has a larger strength for $\pi^+$ (dashed
line) than for $\pi^0$ (solid line), which has a strength larger
than that for $\pi^-$ (dotted line), and this behavior is reversed
at high pion energies. Figure~\ref{delta} shows the mass
distributions of $\Delta$ resonances at rest in asymmetric nuclear
matter as functions of mass. One sees that it is similar to that in
free space (solid line) as a result of the cancelation between the
pion in-medium effects, which enhance the strength at low masses,
and the Pauli-blocking of the nucleon from delta decay, which
reduces the strength at low masses. This is consistent with observed
similar energy dependence of the photo-proton and photo-nucleus
absorption cross sections around the $\Delta$
resonance~\cite{vanPee:2007tw}. Furthermore, whereas the strength
around the peak of the $\Delta$ resonance mass distribution
decreases with increasing charge of $\Delta$ resonance due to
nonzero isospin asymmetry of the nuclear medium, that near the
threshold increases with increasing $\Delta$ resonance charge. We
note that the pion-nucleon $s$-wave and $p$-wave interactions have
opposite effects on the in-medium properties of pions and delta
resonances. While the pion-nucleon $s$-wave interaction increases
the $\pi^-$ mass and reduces the $\pi^+$ mass, the pion-nucleon
$p$-wave interaction softens the dispersion relation of $\pi^-$ more
than that of $\pi^+$. As a result, including only medium effects due
to the pion-nucleon $p$-wave interaction would lead to an opposite
result, i.e., for low momentum pions the spectral function at low
energies has a larger strength for $\pi^-$ than for $\pi^+$,
although the strength around the peak of the $\Delta$ resonance mass
distribution still decreases with increasing charge of $\Delta$
resonance.

\section{Charged pion ratio in hot dense asymmetric nuclear matter}\label{pionratio}

To see the isospin-dependent pion in-medium effects on the
$\pi^-/\pi^+$ ratio in heavy ion collisions, we use a thermal model
which assumes that pions are in thermal equilibrium with nucleons
and $\Delta$ resonances~\cite{bertsch}. The density of particle
species $i$ is then given by
\begin{eqnarray}\label{density}
\rho_i^{} \approx g_i^{}\int \frac{d^3{\bf k}}{(2\pi)^3} d\omega^{n_i^{}} S_i(\omega,k)
\frac{1}{z_i^{-1}e^{(\omega - B_i\mu_B^{}-Q_i\mu_Q^{} )/T} \pm 1}.\notag\\
\end{eqnarray}
In the above, $g_i^{}$, $B_i$, $Q_i$, and $z_i^{}$ are the degeneracy, baryon number,
charge, and fugacity of the particle. The exponent $n_i^{}$ is $2$
for pions and $1$ for nucleons and $\Delta$ resonances. For pions,
we use the spectral functions $S_\pi^{m_t}(\omega,k)$ calculated
above. For the spectral functions of $\Delta$ resonances, we neglect
their momentum dependence and thus replace the integration over
energy $\omega$ by that over mass $M$. The $\omega$ in the
Fermi-Dirac distribution for $\Delta$ resonances is then simply
$\omega=M+k^2/2M+U_\Delta^{m_T}$. For nucleons, their spectral
functions are taken to be delta functions if we neglect the
imaginary part of their self-energies, i.e., $S_N^{m_\tau}(\omega,k)
= \delta ( \omega^{} - m_N^{} - k^2/2m_N^{} - U_N^{m_\tau} ).$

According to studies based on transport
models~\cite{xiong93,bali02,Xiao09}, the total number of pions and
$\Delta$ resonances in heavy ion collisions reaches a maximum value
when the colliding matter achieves the maximum density, and remains
essentially constant during the expansion of the matter. For Au+Au
collisions at the beam energy of $0.4~{\rm AGeV}$, for which the
$\pi^-/\pi^+$ ratio has been measured by the FOPI Collaboration at
GSI~\cite{FOPI}, the IBUU transport model gives a maximum density
that is about twice the normal nuclear matter density and is
insensitive to the stiffness of the nuclear symmetry energy, as it
is mainly determined by the isoscalar part of the nuclear equation
of state~\cite{Xiao09}. We thus use this density in the thermal
model. The temperature in the thermal model is determined by fitting
the measured pion to nucleon ratio, which is about $0.014$ including
pions and nucleons from decays of $\Delta$ resonances~\cite{FOPI},
without medium effects and with unity fugacity parameters for all
particles, and the value is $T \simeq 43.6$~MeV. The assumption that
pions and $\Delta$ resonances are in chemical equilibrium is
consistent with the short chemical equilibration times estimated
from the pion and $\Delta$ resonance production rates. The isospin
asymmetry of the hadronic matter is then taken to be $\delta_{\rm
like} \simeq 0.080$, $0.106$, and $0.143$, corresponding to net
charge densities of $0.920\rho_0^{}$, $0.894\rho_0^{}$ and
$0.857\rho_0^{}$, for the three symmetry energies given by $x=0$,
$0.5$, and $1$, respectively, in order to reproduce the
$\pi^-/\pi^+$ ratios of $2.20$, $2.40$, and $2.60$ predicted by the
IBUU transport model of Ref.~\cite{Xiao09} using corresponding
symmetry energies without pion in-medium effects. Since medium
effects enhance the pion and $\Delta$ resonance densities, to
maintain same pion to nucleon ratio as the measured one requires
fugacity parameters for pions and $\Delta$ resonances to be less
than one. Also, the pion in-medium effects have been shown to affect
only slightly the pion and the $\Delta$ resonance
abundance~\cite{xiong93}, indicating that both pions and $\Delta$
resonances are out of chemical equilibrium with nucleons when medium
effects are included, as expected from estimated increasing pion and
$\Delta$ resonance chemical equilibration times as a result of
medium effects. Because of the small number of pions (about 0.3\%)
and $\Delta$ (about 1.1\%) resonances in the matter, the density,
temperature, and net charge density of the hadronic matter are
expected to remain unchanged when pion in-medium effects are
introduced. They lead to, however, a slight reduction of the isospin
asymmetry to $\delta_{\rm like} \simeq 0.073$, $0.098$, and $0.135$
for the three symmetry energies, given by $x=0$, $0.5$ and $1$,
respectively. We note that with the fugacity of nucleons kept at
$z_N=1$, the fugacity parameters of about $z_\pi^{}=0.0855$ and
$z_\Delta^{}=0.459$ are needed for the symmetry energy given by
$x=1$ to keep the total number of pions as well as that of $\Delta$
resonances the same as in the case without pion in-medium effects,
and that the required values for the fugacity parameters are similar
for the other two symmetry energies considered here.

\begin{figure}[t]
\centerline{\includegraphics[width=3.2in,height=3.2in,angle=0]{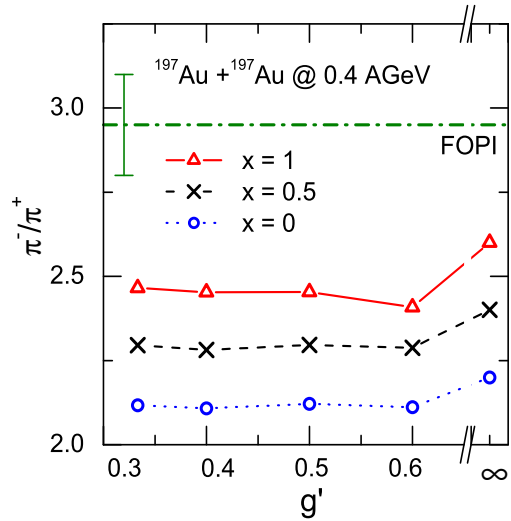}}
\caption{(Color online) The $\pi^-/\pi^+$ ratio in Au+Au collisions
at the beam energy of $0.4~{\rm AGeV}$ for different values of
nuclear symmetry energy parameter ($x=0$, $0.5$, and $1$) and the
Migdal parameter $g^\prime=1/3$, $0.4$, $0.5$, and $0.6$. Results
for $g^\prime=\infty$ correspond to the case without the pion
in-medium effects.} \label{ratio}
\end{figure}

Results on the $\pi^-/\pi^+$ ratio in Au+Au collisions at the beam
energy of $0.4~{\rm AGeV}$ are shown in Fig.~\ref{ratio}. With the
value $g^\prime = 1/3$ for the Migdal parameter, values for the
$\pi^-/\pi^+$ ratio are $2.12$, $2.30$, and $2.47$ for the three
symmetry energy parameters $x=0$, $0.5$, and $1$, respectively,
which are slightly smaller than corresponding values for the case
without including the pion in-medium effects as shown by those for
$g^\prime = \infty$ in Fig.~\ref{ratio}. The measured $\pi^-/\pi^+$
ratio of about $3$ by the FOPI Collaboration, shown in
Fig.~\ref{ratio} by the dash-dotted line together with the error
bar, which without the inclusion of the pion in-medium effects in
the IBUU model favors a nuclear symmetry energy softer than the one
given by $x=1$, thus requires an even softer one after including the
isospin-dependent pion in-medium effects. Figure~\ref{ratio} further
shows results obtained with larger values of $g^\prime=0.4$, $0.5$
and $0.6$ for the Migdal parameter. It is seen that the
isospin-dependent pion in-medium effects in these cases are not much
different from the case of $g^\prime = 1/3$, indicating that the
pion abundance in hot dense asymmetric nuclear matter is affected
more by the $s$-wave, which does not depend on $g^\prime$, than by
the $p$-wave interaction between pion and nucleon.

\section{Summary}\label{summary}

In summary, we have studied the dependence of the pion spectral
function in asymmetric nuclear matter on the charge of the pion by
using results from the chiral perturbation theory for the
pion-nucleon $s$-wave interaction and from the $\Delta$-hole model
for the pion-nucleon $p$-wave interaction. Because of increasing
$\pi^-$ and decreasing $\pi^+$ in-medium masses due to the
pion-nucleon $s$-wave interaction in neutron-rich matter, the
strength of $\pi^+$ spectral function at low energies is somewhat
larger than that of $\pi^-$ spectral function, and the strength
around the peak of the $\Delta$ resonance mass distribution
decreases while that near the threshold increases with increasing
charge of the $\Delta$ resonance. In a thermal model that assumes
that nucleons, pions, and $\Delta$ resonances produced in heavy ion
collisions are in thermal but not chemical equilibrium, with the
latter needed to maintain the final pion to nucleon ratio, the
$\pi^-/\pi^+$ ratio is slightly reduced in comparison with the case
without pion in-medium effects. Taking into consideration of the
isospin-dependent pion in-medium effects in the transport model thus
would have some, albeit not very significant, effects on the
extraction of the nuclear symmetry energy from measured
$\pi^-/\pi^+$ ratio. On the other hand, the predicted $\pi^- /
\pi^+$ ratio for a given parametrization of the symmetry energy can
be quite different in different transport models.  For example, it
was shown in Ref.~\cite{Ferrini:2005jw} that the $\pi^-/\pi^+$ ratio
in hot dense nuclear matter increases with increasing stiffness of
the nuclear symmetry energy at high densities if the latter is
parameterized using the relativistic mean-field model, opposite to
the results from the IBUU model~\cite{Xiao09} which shows a
decreasing $\pi^-/\pi^+$ ratio with increasing stiffness of the
nuclear symmetry energy.  Also, it was recently shown in an improved
isospin-dependent quantum molecular dynamic (ImIQMD) model that the
measured $\pi^-/\pi^+$ ratio from the FOPI collaboration is
consistent with a nuclear symmetry energy that is even stiffer than
the one corresponding to the symmetry energy parameter $x=0$ in the
MDI interaction~\cite{Zhao10}, contradictory to the conclusion from
the IBUU model~\cite{Xiao09} that a symmetry energy softer than
$x=1$ is needed to reproduce the measured $\pi^-/\pi^+$ ratio.
Further theoretical work is thus needed to understand the relation
between the $\pi^-/\pi^+$ ratio and the behavior of the nuclear
symmetry energy at high densities in the transport model description
of heavy ion collisions.

\section*{Acknowledgements}

This work was supported in part by the US National Science
Foundation under Grant No. PHY-0758115 and the Welch Foundation
under Grant No. A-1358.

\end{document}